\documentclass{article}
\usepackage{spconf,amsmath,graphicx, cite, multirow, stfloats, booktabs, amsmath, amsfonts, amssymb, bm}
\usepackage{xcolor}
\usepackage{arydshln}
\usepackage{threeparttable}
\usepackage[colorlinks,linkcolor=red]{hyperref}

 \title{Anomalous Sound Detection using Spectral-Temporal Information Fusion}

\name{Youde Liu$^1$, Jian Guan$^{1*}$\thanks{*Corresponding Author}, Qiaoxi Zhu$^2$, Wenwu Wang$^3$
\thanks{This work was partly supported by the Natural Science Foundation of Heilongjiang Province under Grant No. YQ2020F010, 
and a Newton Institutional Links Award from the British Council with Grant No. 623805725.
}}
\address{
  $^1$Group of Intelligent Signal Processing, College of Computer Science and Technology\\Harbin Engineering University,  China\\
  $^2$Centre for Audio, Acoustics and Vibration, University of Technology Sydney,  Australia\\
  $^3$Centre for Vision Speech and Signal Processing, University of Surrey,  UK}
\begin{document}
%
\maketitle
\begin{abstract}
Unsupervised anomalous sound detection aims to detect unknown abnormal sounds of machines from normal sounds. However, the state-of-the-art approaches are not always stable and perform dramatically differently even for machines of the same type, making it impractical for general applications. This paper proposes a spectral-temporal fusion based self-supervised method to model the feature of the normal sound, which improves the stability and performance consistency in detection of anomalous sounds from individual machines, even of the same type. Experiments on the DCASE 2020 Challenge Task 2 dataset show that the proposed method achieved 81.39\%,  83.48\%, 98.22\% and 98.83\% in terms of the minimum AUC (worst-case detection performance amongst individuals) in four types of real machines (fan, pump, slider and valve), respectively, giving 31.79\%, 17.78\%, 10.42\% and 21.13\% improvement compared to the state-of-the-art method, i.e., Glow\_Aff. Moreover, the proposed method has improved AUC (average performance of individuals) for all the types of machines in the dataset. The source codes are available at \url{https://github.com/liuyoude/STgram_MFN}
\end{abstract}
\section{Introduction}
\label{sec:intro}
Anomalous sound detection (ASD) aims to automatically identify whether a target object (e.g., a machine or a device) is normal or abnormal from the sound emitted. Collecting  anomalous sound data is not a trivial task due to their diversity and scarcity in the real world. Therefore, normal sounds are often used to learn the features of the normal sound, and these learnt features are then used to distinguish the normal and abnormal sound\cite{oh2018residual, park2018fast, Koizumi_DCASE2020_01}.

Conventional ASD systems have used autoencoder (AE), such as interpolation deep neural network (IDNN) \cite{suefusa2020anomalous} and ID-conditioned autoencoder \cite{Kapka2020}. They learn the feature of normal sounds by minimizing the reconstruction error and using the reconstruction error as the score to detect the anomalies. However, since the training procedure does not involve the anomalous sound, the effectiveness of such model could be limited. if the trained feature also fits with the anomalous sound \cite{koizumi2018unsupervised}.

To better model the normal sound feature, the self-supervised classification approach has been proposed by using the metadata, i.e., machine type and machine identity (ID) in addition to condition (normal/anomaly), accompanying the audio files \cite{giri2020self}, and it performs better than the AE-based unsupervised methods. However, this method is not always stable and performs differently even for machines of the same type \cite{dohi2021flow}. As a solution, the flow-based self-supervised density estimation (i.e., Glow\_Aff) was proposed in \cite{dohi2021flow}, using normalizing flow such as generative flow (Glow) \cite{kingma2018glow} or masked autoregressive flow (MAF) \cite{papamakarios2017masked}. This method improves the detection performance on one machine ID by introducing an auxiliary task to distinguish the sound data of that machine ID (target data) from sound data of other machine IDs with the same machine type (outlier data). Thus, this approach requires different training models for different machine IDs of each machine type, which is not desired for general applications. In addition, the detection stability of this approach on sounds of individual machines of the same type is still limited.

To develop a general method with stability and avoid specialized training for each machine ID, we study empirically the effectiveness of features for ASD, such as the log-Mel spectrogram which has been widely used as input feature in ASD \cite{suefusa2020anomalous, Kapka2020, giri2020self, dohi2021flow}. This feature was designed based on human auditory perception, using mel filter bank to capture the information in various frequencies \cite{1163420}. However, it might filter out high-frequency components of anomaly sound, where distinct features may exist. Thus, the log-Mel spectrogram feature may not be able to fully distinguish the normal and anomalous sounds, resulting in unstable performance when used with the self-supervised approaches. There is potential to use temporal information to complement the log-Mel spectrogram, which is recently studied in audio pattern recognition \cite{9229505}. However, the incorporation of temporal information has not been reported for ASD.

In this paper, a spectral-temporal feature, STgram, is proposed as the input feature for self-supervised classification approach by fusing the log-Mel spectrogram (Sgram) and the temporal feature (Tgram). Here, the temporal feature is extracted from the raw wave by a proposed CNN-based network (TgramNet) to compensate for the anomalous information unavailable from the log-Mel spectrogram. The spectral-temporal feature is then fed into the classifier, i.e., MobileFaceNet (MFN) \cite{chen2018mobilefacenets}, to learn the delicate feature representation of normal sounds for the detection of anomalies.

The proposed method is evaluated by experiments on the DCASE 2020 Challenge Task 2 dataset\cite{Koizumi_DCASE2020_01}, as compared with the state-of-the-art methods, over six types of machine sound. The proposed method improves AUC and pAUC performance and largely increases performance stability observed from the mAUC metric. Furthermore, the ablation study presents the comparison with potential alternatives using temporal only or spectral-temporal input feature, to show that the proposed method provides an effective way of utilizing both the spectral and temporal features for anomalous sound detection.
\begin{figure*}[!ht]
\begin{minipage}[b]{.82\linewidth}
 \centering
 \centerline{\includegraphics[width=0.85\textwidth]{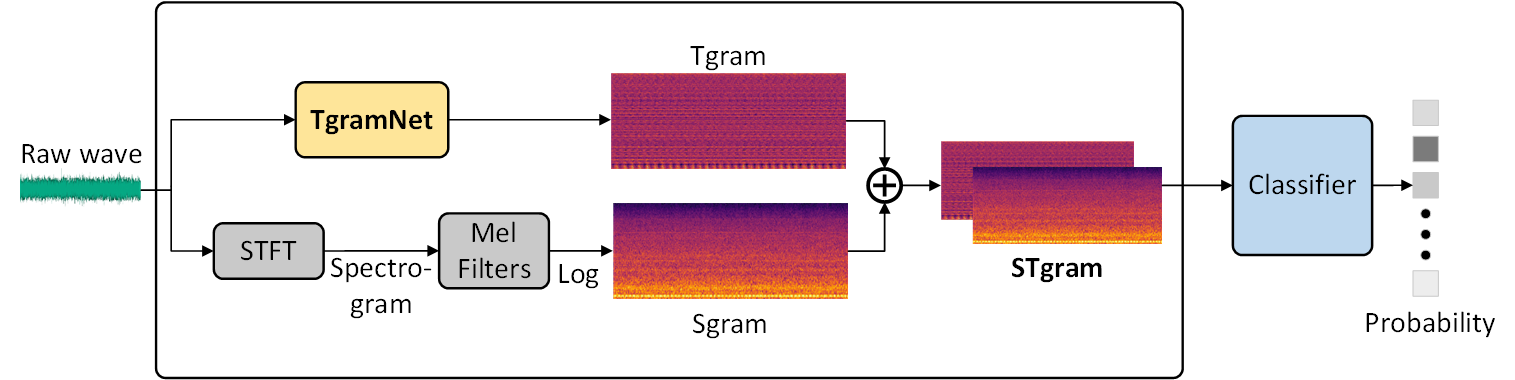}}
  \centerline{(a)}\medskip
\end{minipage}
\hspace{0.1cm}
\begin{minipage}[b]{.14\linewidth}
 \centering
 \centerline{\includegraphics[width=0.8\textwidth]{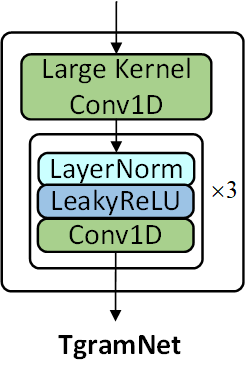}}
 \centerline{(b)}\medskip
\end{minipage}\caption{The framework of the proposed method for anomalous sound detection. (a) The spectral-temporal feature is extracted from the raw wave through a CNN-based network (TgramNet) for the temporal feature and the log-Mel spectrogram for the frequency feature. $\bm  \oplus$ denotes concatenation operation for feature fusion. The spectral-temporal feature is then fed into the classifier for anomalous sound detection.  (b) Details in TgramNet.}
\label{fig:1}
\vspace{-0.4cm}
\end{figure*}

\section{Proposed Method}
\label{sec:propose}

This section presents the proposed method for anomalous sound detection in detail. The overall framework of the proposed method is given in Fig.~\ref{fig:1}.

\subsection{Spectral-temporal feature fusion}

Let $\bm x \in \mathbb{R}^{1 \times L}$ be the input single-channel audio signal with the length $L$. The log-Mel spectrogram of $\bm x $ is $\bm F_S \in \mathbb{R}^{M \times N}$, where $M$ denotes the dimension of the spectrum feature (i.e., number of Mel bins) and $N$ is the number of time frames. {The log-Mel spectrogram can be obtained as follows:}
\begin{equation}
\label{eq:1}
    \bm F_S = log(\mathcal{\bm{W}}_M \cdot \|STFT(\bm x)\|^2),
\end{equation}
where $\mathcal{\bm{W}}_M \in \mathbb{R}^{M \times B}$ represents the Mel-filter banks and $B$ is the number of frequency bins of the spectrogram, obtained by short-time Fourier transform (STFT).

To compensate for the missing anomaly information from the log-Mel spectrogram, we apply a CNN-based network (TgramNet) to extract the temporal feature from the sound signal $\bm x$. The architecture of TgramNet is shown in Fig.~\ref{fig:1}(b) and Table~\ref{tab:1}. Firstly, a large kernel 1D convolution is used, with channel number, kernel size and stride set the same as the number of Mel bins, window size and hop length for the log-Mel spectrogram. Then, three CNN blocks are applied, and each block consists of a layer normalization \cite{ba2016layer}, a leaky ReLU activation function, and a 1D convolution with a smaller kernel size. Note that, the CNN blocks do not change the dimension of the output temporal feature. Thus, the resultant temporal feature, Tgram, is
\begin{equation}
\label{eq:2}
\bm F_T =  TN(\bm x),
\end{equation}
where $TN(\cdot)$ represents the TgramNet for feature extraction from time domain, and $\bm {F}_T \in \mathbb{R}^{M \times N}$ has the same dimension as $\bm F_S$. 
\begin{table}[!t]
\centering
\setlength{\belowcaptionskip}{1pt}
\setlength{\abovecaptionskip}{1pt}
\caption{The architecture of TgramNet.}
\begin{threeparttable}
\begin{tabular}{ccccc:c}
\hline
Operation & c & k & s & p & n                  \\ \hline
Conv1D    & $M$ & $W$ & $H$ & $W/2$  & $\times 1$                  \\ \hline

LayerNorm & - & - & - & - &          \multirow{3}{*}{$\times 3$}\\
LeakyReLU & - & - & - & - &                  \\ 
Conv1D    & $M$ & 3 & 1 & 1 &  \\\hline
\end{tabular}
\begin{tablenotes}
       \footnotesize
       \item[*] Here, c, k, s, p, n represent number of channels, kernel size, stride, padding size and number of layers, respectively. $M$ is the number of Mel bins.  $W$ and $H$ are the window size and hop length of STFT, respectively.
     \end{tablenotes}
\end{threeparttable}
\label{tab:1}
\vspace{-0.3cm}
\end{table}

Finally, the spectral-temporal fusion feature STgram $\bm F_{ST} \in \mathbb{R}^{2 \times M \times N}$ is obtained using a simple fusion strategy by concatenating the log-Mel spectrogram  $\bm F_S$ and  Tgram $\bm F_T$, that
\begin{equation}
\label{eq:3}
\bm F_{ST} =  Concat(\bm F_S, \bm F_T),
\end{equation}
where
$ Concat(\cdot)$ denotes the concatenation operation. 

\subsection{Self-supervised classification}
We adopt a self-supervised classification strategy following \cite{giri2020self}, where the metadata (i.e., machine IDs) accompanying the audio feature (i.e., STgram) are used to learn feature representations of normal sound, resulting in the better ability of the model in distinguishing the normal and abnormal sound. Specifically, we choose MobileFaceNet (MFN) \cite{chen2018mobilefacenets} as the baseline classifier to learn the delicate representation of normal sounds. The whole method is abbreviated as STgram-MFN. For better sensitivity to the anomalies, STgram-MFN applies ArcFace \cite{deng2019arcface}, rather than the cross-entropy error (CEE), as the loss function which helps increase the distance between classes and decrease the distance within classes.
\begin{table*}[htbp]
\scriptsize
    \centering
	\setlength{\belowcaptionskip}{1pt}
	\setlength{\abovecaptionskip}{1pt}
	\caption{Performance comparison in terms of AUC (\%) and pAUC (\%) for different types of machines.}
		\begin{tabular}{ccccccccccccccccccccccc}
			\toprule
			\multicolumn{2}{c}{\multirow{2}[2]{*}{Methods}} &\multicolumn{2}{c}{Fan}&\multicolumn{2}{c}{Pump}&\multicolumn{2}{c}{Slider}&\multicolumn{2}{c}{Valve}&\multicolumn{2}{c}{ToyCar}&\multicolumn{2}{c}{ToyConveyor}&\multicolumn{2}{c}{Average} \\
			\cmidrule(r){3-4} \cmidrule(r){5-6} \cmidrule(r){7-8} \cmidrule(r){9-10} \cmidrule(r){11-12} \cmidrule(r){13-14} \cmidrule(r){15-16} 
			\multicolumn{2}{c}{} & {AUC} & {pAUC} & {AUC} & {pAUC} & {AUC} & {pAUC} & {AUC} & {pAUC}& {AUC} & {pAUC} & {AUC} & {pAUC} & {AUC} & {pAUC} \\
			\midrule
			\multicolumn{2}{c}{IDNN\cite{suefusa2020anomalous}} 
			& 67.71  & 52.90  & 73.76  & 61.07  & 86.45  & 67.58  & 84.09  & 64.94  & 78.69  & 69.22  & 71.07  & 59.70  & 76.96  & 62.57 \\
			\multicolumn{2}{c}{MobileNetV2\cite{giri2020self}} 
			& 80.19  & 74.40  & 82.53  & 76.50  & 95.27  & 85.22  & 88.65  & 87.98  & 87.66  & 85.92  & 69.71  & 56.43  & 84.34  & 77.74 \\
		\multicolumn{2}{c}{Glow\_Aff\cite{dohi2021flow}}
		& 74.90  & 65.30  & 83.40  & 73.80  & 94.60  & 82.80  & 91.40  & 75.00  & 92.20  & 84.10 
		& 71.50  & 59.00 
		& 85.20  &73.90 \\
		\multicolumn{2}{c}{\textbf{STgram-MFN(CEE)}} & 91.30  & 86.73  & 91.25  & 81.69  & 99.36  & 96.84  & 94.44  & 91.58  & 88.80  & 87.38  & 72.93  & \textbf{63.62}  & 89.68  & 84.64 \\
		\multicolumn{2}{c}{\textbf{STgram-MFN(ArcFace)}} & \textbf{94.04}  & \textbf{88.97}  & \textbf{91.94}  & \textbf{81.75}  & \textbf{99.55}  & \textbf{97.61}  & \textbf{99.64}  & \textbf{98.44}  & \textbf{94.44}  & \textbf{87.68}  & \textbf{74.57}  & 63.60  & \textbf{92.36}  & \textbf{86.34} \\
			\bottomrule
			\bottomrule
			\end{tabular}
	\label{tab:2}
\end{table*}

\section{Experiments and Results}
\label{sec:exper}
\subsection{Experimental setup}
\label{ssec:setup}
\textbf{\textit{Dataset}} 
We evaluate our method using DCASE 2020 challenge Task2 development and additional dataset \cite{Koizumi_DCASE2020_01}, which consists of part of MIMII \cite{Purohit_DCASE2019_01} and ToyADMOS dataset \cite{Koizumi_WASPAA2019_01}. The MIMII dataset has four machine types (i.e., Fan, Pump, Slider and Valve), and each type includes seven different machines. The ToyADMOS dataset has two machine types (i.e., ToyCar and ToyConveyor), including seven and six different machines, respectively. Here, machine ID is used to identify different machines with the same machine type. In the experiments, the training data (normal sound) from the development and additional dataset of Task 2 \cite{Koizumi_DCASE2020_01} are used as the training set, and the test data (normal and anomaly sound) of the development dataset is employed for evaluation.

\noindent \textbf{\textit{Evaluation metrics}} The performance is evaluated with the area under the receiver operating characteristic (ROC) curve (AUC) and the partial-AUC (pAUC), following \cite{suefusa2020anomalous, giri2020self, dohi2021flow}, where pAUC is calculated as the AUC over a low false-positive-rate (FPR) range $[0, p]$ and $p = 0.1$ as in \cite{Koizumi_DCASE2020_01}. In addition, the minimum AUC (mAUC) is taken to represent the worst detection performance achieved among individual machines of the same machine type, following \cite{dohi2021flow}. 

\noindent \textbf{\textit{Implementation}} We train our proposed STgram-MFN on the training set of raw wave audio signals with a length of around 10 seconds, where one model is trained for all machine types. The frame size is 1024 samples with an overlapping 50\% for the log-Mel spectrogram, and the number of Mel filter banks is 128 (i.e., $W=1024$, $H=512$ and $M=128$). Accordingly, the obtained Sgram $\bm F_S$ and Tgram $\bm F_T$ have a dimension of $128 \times 313$. Adam optimizer \cite{kingma2014adam} is employed for model training with a learning rate of  0.0001, and the cosine annealing strategy is adopted for learning rate decay. The model is trained with 200 epochs, and the batch size is 128. The margin and scale parameters of ArcFace \cite{deng2019arcface} are 0.7 and 30, respectively. The negative log probability is used as the anomaly score for detection.
\vspace{-0.3cm}
\begin{table}[h]
\centering
\caption{ Performance comparison in terms of mAUC (\%).}
\resizebox{\linewidth}{18mm}{
\begin{tabular}{cccccc}
\toprule
Methods & IDNN\cite{suefusa2020anomalous}
& \begin{tabular}[c]{@{}c@{}}Mobile\\NetV2\cite{giri2020self}\end{tabular} 
& \begin{tabular}[c]{@{}c@{}}Glow\_Aff\cite{dohi2021flow}\end{tabular} 
& \begin{tabular}[c]{@{}c@{}c@{}}\textbf{STgram}\\\textbf{-MFN}\\{\textbf{(CEE)}}\end{tabular} 
& \begin{tabular}[c]{@{}c@{}c@{}}\textbf{STgram}\\\textbf{-MFN}\\{\textbf{(ArcFace)}}\end{tabular} \\ 
\midrule
Fan        & 56.56  & 50.40       & 49.60 & 79.80           & \textbf{81.39}                      \\
Pump       & 61.86  & 52.90       & 65.70 & 79.79           & \textbf{83.48}                      \\
Slider     & 74.22  & 82.80       & 87.80 & \textbf{98.39}  & {98.22}                             \\
Valve      & 66.83  & 67.90       & 77.70 & 79.12           & \textbf{98.83}                      \\
ToyCar     & 64.41  & 55.70       & 80.10 & 61.91           & \textbf{83.07}                      \\
ToyConveyor& 62.89  & 48.70       & 61.00 & 57.25           & \textbf{64.16}                      \\
\midrule
Average    & 64.46  & 59.73       & 70.32 & 76.04           & \textbf{84.86}                      \\ 
\bottomrule
\bottomrule
\end{tabular}
}
\label{tab:3}
\end{table}

\vspace{-0.4cm}
\subsection{Performance comparison}
\label{ssec:compare}
Table \ref{tab:2} shows the comparison of the proposed STgram-MFN with other state-of-the-art methods, IDNN \cite{suefusa2020anomalous}, MobileNetV2 \cite{giri2020self}, and Glow\_Aff \cite{dohi2021flow}. IDNN is the AE-based method, MoblieNetV2 is the self-supervised classification method, and Glow\_Aff is the flow-based self-supervised method. Regarding STgram-MFN, CEE and ArcFace loss are adopted for model training, respectively, denoted as STgram-MFN(CEE) and STgram-MFN(ArcFace). It is shown that the proposed method significantly improves the ASD performance, specifically 7.16\% improvement on AUC and 8.6\% improvement on pAUC (averaged over all the six machine types), compared with the best performance achieved by other methods in the literature.

Table \ref{tab:3} shows the mAUC results to reflect the worst detection performance achieved among individual machines of the same type. The instability of the previous methods can be observed that MobileNetV2 outperforms IDNN in Table \ref{tab:2}, but its average mAUC (59.73\%) is worse than IDNN (64.46\%) in Table \ref{tab:3}. Amongst all the methods, STgram-MFNs achieves the best result with a greater mAUC improvement compared to Glow\_Aff. Specifically, STgram-MFN (ArcFace) achieves the average mAUC of 84.86\%, outperforming Glow\_Aff by 14.54\%.
\begin{table*}[htbp]
\scriptsize
    \centering
	\caption{
	Performance comparison for different input features.}
		\begin{tabular}{cccccccccccccccccc}
			\toprule
			\multicolumn{1}{c}{\multirow{2}[2]{*}{Methods}} &\multicolumn{2}{c}{\ \ \ \ \ \ LogMel-MFN\ \ \ \ \ \ }&\multicolumn{2}{c}{\ \ \ \ \ \ Tgram-MFN\ \ \ \ \ \ }&\multicolumn{2}{c}{\ \ \ \ \ \ Spec-MFN\ \ \ \ \ \ }&\multicolumn{2}{c}{\begin{tabular}[c]{@{}c@{}}\textbf{STgram-MFN(CEE)}\end{tabular}}&\multicolumn{2}{c}{\begin{tabular}[c]{@{}c@{}}\textbf{STgram-MFN(ArcFace)}\end{tabular}} \\
            \cmidrule(r){2-3} \cmidrule(r){4-5} \cmidrule(r){6-7} \cmidrule(r){8-9} \cmidrule(r){10-11}
			\multicolumn{1}{c}{} & {AUC} & {mAUC} & {AUC} & {mAUC} & {AUC} & {mAUC} & {AUC} & {mAUC}& {AUC} & {mAUC}& \\
			\midrule
Fan         & 82.36 & 53.75 & 89.47 &\textbf{83.85}& 80.36 & 49.75 & 91.30 &79.80 & \textbf{94.04} &81.39\\
Pump        & 87.74 & 67.62 & 89.13 &82.60 & 83.73 & 65.92& 91.25 &79.79 & \textbf{91.94} &\textbf{83.48}\\
Slider      & 99.08 & 98.07 & 71.64 &27.45 & 93.62 &88.62 & 99.36 &\textbf{98.39} & \textbf{99.55} & 98.22\\
Valve       & 89.91 & 65.88 & 87.41 &74.32 & 86.46 &77.77& 94.44 &79.12 & \textbf{99.64} & \textbf{98.83}\\
ToyCar      & 88.73 & 66.32 & 60.72 &49.89 & 64.07 &47.22& 88.80 &61.91 & \textbf{94.44} & \textbf{83.07}\\
ToyConveyor & \textbf{78.17} &\textbf{67.79} & 52.70 & 46.71 & 54.47 & 51.85 & 72.93&57.25 & 74.57 & 64.16\\ 
\midrule
Average     & 87.67&69.91& 75.18&60.80& 77.12&63.52&  89.68&76.04& \textbf{92.36}&\textbf{84.86}\\ 
			\bottomrule
			\bottomrule
			\end{tabular}
	\label{tab:4}
\end{table*}

\begin{figure}[!ht]
\begin{minipage}[b]{.95\linewidth}
 \centering
 \centerline{\includegraphics[width=0.80\textwidth]{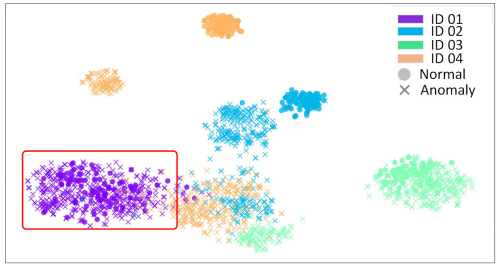}}
  \centerline{(a) log-Mel spectrogram}\medskip
\end{minipage}
\begin{minipage}[b]{.95\linewidth}
 \centering
 \centerline{\includegraphics[width=0.80\textwidth]{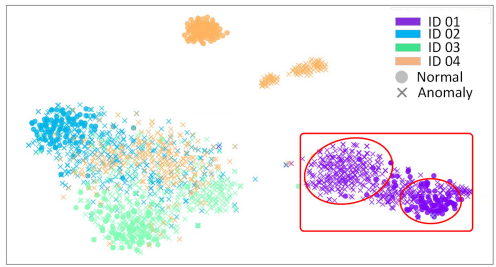}}
 \centerline{(b) Tgram}\medskip
\end{minipage}
\caption{The t-SNE visualization of log-Mel spectrogram and Tgram for MFN classifier on the test dataset for the machine type Fan. Different color represents different machine ID. The ``$\bullet$'' and ``$\times$'' denote normal and anomalous classes, respectively. The normal and anomaly cluster for ``ID 01'' is highlighted by contours in red.}
\label{fig:2}
\end{figure}

\subsection{Ablation study}
\label{ssec:ablation}
To show the effectiveness of STgram, we conducted an ablation study using log-Mel spectrogram, Tgram, spectrogram and STgram, respectively, as the input features of MFN, and the results are presented as LogMel-MFN, Tgram-MFN, Spec-MFN and STgram-MFN in Table \ref{tab:4}. As ArcFace needs to adjust parameters for different methods, we only use CEE as the loss function in LogMel-MFN, Tgram-MFN and Spec-MFN for a fair comparison. 

Table \ref{tab:4} shows that LogMel-MFN has a much smaller mAUC than AUC on Fan, Pump, Valve, and ToyCar, reflecting inconsistent performance for different machines even of the same type. Tgram-MFN performs better in terms of mAUC on Fan, Pump, and Valve than LogMel-MFN. However, log-Mel spectrogram and Tgram are complementary, as illustrated by t-distributed stochastic neighbor embedding (t-SNE) cluster visualization of the latent features of log-Mel spectrogram and Tgram in  Fig. \ref{fig:2}. For example, it is clear that the anomalous and normal latent features of machine ``ID 01" are overlapping in terms of the log-Mel spectrogram, while they are more distinguishable by Tgram. This finding shows that the log-Mel spectrogram may filter out useful information about anomalies.

However, Tgram-MFN does not perform well in general, since it may suffer from noise contained in the temporal information. We also evaluated spectrogram without Mel filtering as the input feature (Spec-MFN in Table \ref{tab:4}). It can be observed that Spec-MFN performs better on Slider and Valve but worse on Fan and Pump than Tgram-MFN. Compared to the above potential alternatives, the proposed input feature is much more effective for anomalous sound detection. STgram-MFN(CEE) and STgram-MFN(ArcFace) achieve much higher mAUC (76.04\% and 84.86\%, respectively), as compared with those of LogMel-MFN (69.91\%), Tgram-MFN (60.80\%), and Spec-MFN (63.52\%), suggesting that they offer more stable detection performance.

\section{Conclusion}
\label{sec:conclusion}
In this paper, we have presented a self-supervised anomalous sound detection method, where a spectral-temporal fusion feature from the raw wave is applied, by combining temporal information from a CNN network and spectral information from the log-Mel spectrogram. The proposed method  exploits complementary spectral-temporal information from the normal sound via the fused features, and results in more stable detection performance amongst different machines. The experimental results demonstrated the effectiveness of the proposed method with substantial improvements over the state-of-the-art methods. 
\bibliographystyle{IEEEtran}
\bibliography{refs}

\end{document}